\documentclass[twocolumn, prb]{revtex4}

\usepackage{graphicx}
\usepackage{dcolumn}
\usepackage{bm}

\begin{document}

\title{ On the universality class of the Mott transition in two dimensions}

\author { S. Moukouri$^1$, E. Eidelstein$^2$}

\affiliation{$^1$Racah Institute of Physics, Hebrew University, Jerusalem 91904 
Israel.\\
$^2$Department of Physics, NRCN, P.O. Box 9001, IL Beer-Sheva, 84190 Israel.
} 

\begin{abstract}
We use the two-step density-matrix renormalization 
group method to elucidate the long-standing issue  of the universality
class of the Mott transition in the Hubbard model in two dimensions. 
We studied  a spatially anisotropic two-dimensional Hubbard model with a 
non-perfectly nested Fermi surface at half-filling. We find that unlike 
the pure one-dimensional case where
there is no metallic phase, the quasi one-dimensional model
displays a genuine metal-insulator transition at a finite value of the 
interaction. The critical exponent of the correlation length is found to
be  $\nu \approx 1.0$. This implies that the fermionic Mott transition, 
 belongs to the universality class of the 2D Ising  model. 
The Mott insulator is the 'ordered' phase whose order parameter is
given by the density of singly occupied sites minus that of holes
and doubly occupied sites. 
\end{abstract}

\maketitle

\section{Introduction}
\label{intro}

 In the studies of the Mott transition \cite{mott,imada-RMP} in the ground 
state of the Hubbard model \cite{hubbard}, there are well controlled results 
in the pure one-dimensional (1D) case \cite{lieb-wu} and in the limit of 
infinite dimensions \cite{metzner-vollhardt,georges-kotliar,georges-RMP} only. 
In 1D, there is no metallic phase, the Mott gap opens as soon as the 
interaction $U>0$. In infinite 
dimensions, the dynamical mean-field theory which is exact predicts a Mott 
transition at the critical coupling, $U_c \approx W$, $W$ is the band width. 
However, the transition has mean-field critical exponents. This anomaly is 
due to the local nature of the infinite dimensional solution. Hence, the 
one-dimensional and the infinite dimensional solutions may not be directly 
applicable to experiments. Studies of the Mott transition in the Hubbard 
beyond these special limits of one dimension and infinite dimension are thus 
of crucial importance.

 For more than a decade, a great deal of effort has been devoted to applying
quantum cluster theories \cite{moukouri-jarrell,zhang-imada, parcollet,maier-RMP,ohashi,park-haule,potthoff} to
the study of the Mott transition in the Hubbard model in two dimensions (2D). 
Quantum cluster theories include non-local correlations. They 
predict a finite critical value for the interaction at the transition.
This critical value depends on the cluster size. However, when applied 
to a finite dimensional 
model, they are exact only in the limit of infinite cluster size. In quantum 
cluster theories, the effect of the interaction on physical quantities such as 
the single-particle Green's function is restricted to the cluster sites. 
The correlation are fully accounted for distances which are smaller than the 
cluster length, $r \alt L_c$. When $r \agt L_c$, the Green's function has an 
effective mean-field decay. Restricting the effect of the interaction at 
distances $r \alt L_c$ is probably justified away enough from the critical 
point where the correlations are expected to be short-ranged. A consequence of 
this restriction of the correlations to the cluster length is that the 
exponents at the transition are always mean-field like for a fixed cluster
size \cite{maier-RMP}. A systematic finite cluster size analysis is therefore 
necessary for a correct description of the transition. However, most of 
applications of quantum cluster simulations have been done on relatively 
small clusters. These are not enough to reliably predict the low-energy 
physics at the quantum critical point. 
     
Unlike the fermionic model, in the 2D Bose-Hubbard model which displays a 
transition from a superfluid to a Mott insulator, analytical approaches 
\cite{fisher,elstner} and large scale Monte Carlo simulations \cite{capogrosso}
have yielded reliable information about its critical behavior. The transition 
for fixed boson density belongs to the  universality class of the classical
three-dimensional (3D) XY model. This has also been reported on the 2D 
Jaynes-Cummings-Hubbard model \cite{hohenadler}. Unfortunately, for
the fermionic Hubbard model Monte Carlo simulations predict $U_c=0$.
This is because of the nesting induced Slater transition \cite{hirsch,varney}.
In absence of perfect nesting, the Monte Carlo method is hampered by
the sign problem. Large scale simulations are not possible.

Recent interest has been raised by slave rotor analyses 
\cite{florens-georges, senthil}. These analyses suggest that the 
transition in the 2D fermionic Hubbard  model may belong to the 3D XY 
universality class as the bosonic Hubbard model. 
In Ref.\cite{florens-georges,senthil} a slave 
rotor representation of the fermionic operator $c_{i\sigma}=b_if_{i\sigma}$, 
where $b_i$ is a spinless boson and $f_{i\sigma}$ a charge-less spin, was used
to map the Hubbard model to a free spinon Hamiltonian self-consistently coupled
to a bosonic term (or XY term in a spin representation of bosons).  The 
fermionic Mott transition is in this form a transition between condensed 
(Fermi liquid) and non-condensed (Mott insulator) phases of bosons. This
 factorization may be justified in the Mott phase where, because of the Mott 
gap, spin and charge degrees of freedom may be separated. However, as the 
critical point is approached, is the gauge field weak enough to justify the 
decoupling between spin and charge? If not would that modify the critical 
behavior predicted by the slave-rotor approximation? Only a non-biased 
calculation of the Hubbard model can yield the answer.

The slave-rotor prediction is in disagreement with an earlier approximate
mapping \cite{castellani} of the Hubbard model to a generalized 
Blume-Emery-Griffiths model \cite{blume} of the $H_e^3-H_e^4$ mixtures
with an additional term whose effect on the nature of the transition
is not known. In this mapping, doubly occupied and empty sites corresponds 
to $H_e^4$ sites and singly occupied sites to $H_e^3$ sites. This mapping 
suggests instead that the Hubbard model is in the universality class of the 
Ising model. But the extra term which accompanies the Blume-Emery-Griffiths
model could well lead to another universality class.

 In a recent paper \cite{moukouri-eidelstein}, we reported a two-step 
density-matrix renormalization group (DMRG) \cite{moukouri} study of the  Mott 
transition in the ground state of the quasi-one-dimensional (1D) Hubbard 
model at half-filling. We find that in contrast to the pure 1D case for which 
there is no metallic phase, there is an authentic Mott transition in the 
quasi-1D model. However, it is possible to argue that in the quasi-1D 
dimensional Hubbard model 
studied in Ref.\cite{moukouri-eidelstein}, the Fermi surface is perfectly 
nested, thus our analysis which predicts a gapless phase in the weak-coupling 
regime, would miss an exponentially small gap 
$\Delta \propto exp-\frac{2\pi t}{U}$, that would open as a consequence of 
a Slater transition. However, our numerical data did not support the existence 
of such a gap. Arguments supporting a gap opening induced by perfect nesting 
are perturbative: the divergence of the non-interacting susceptibility 
$\chi_0({\bf q})$ at the nesting wave vector leads to that of the interacting 
spin susceptibility, $\chi_s({\bf q}) \propto 1/(1-U\chi_0({\bf q}))$. However,
 the actual susceptibilities and interaction in the expression of 
$\chi_s({\bf q})$ are renormalized. Attempts to compute the renormalized 
susceptibilities and interaction within the self-consistent parquet 
formalism \cite{bickers} lead to intractable equations. Hence, the effect 
of these renormalization effects on the mean-field solution remains an open 
problem.

In this paper, we present a well controlled study of the Mott transition 
in the Hubbard model with a non-perfectly nested Fermi surface beyond the 
special cases of 1D and infinite dimensions. The choice of the non-perfectly
nested Fermi surface precludes the theoretical possibility of a gap induced 
by the Slater anti-ferromagnetism mechanism. The two-step DMRG method is first
checked on the transition between a paramagnetic and an anti-ferromagnetic
ground states in the quasi-1D Heisenberg model with $S=1$. In agreement with
a quantum Monte carlo study \cite{matsumoto}, we find that this transition 
belongs to universality class of the  3D classical Heisenberg model. For the
quasi-1D Hubbard model, we find that, in contrast to the pure one-dimensional 
model, there is a genuine ground-state Mott transition at a finite critical 
value of the interaction. Data analysis of the critical behavior of this model
show that, in agreement with the mapping to the Blume-Emery-Griffiths model
\cite{castellani}, the Mott transition in the 2D Hubbard model belongs to the 
universality class of the 2D Ising model.

\section{Model}
\label{model}

\begin{figure}
\begin{center}
\includegraphics[width=6.cm, height=4cm]{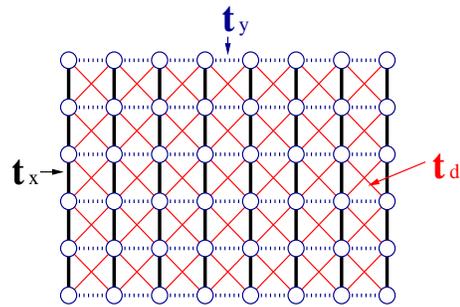}
\end{center}
\caption{ The anisotropic frustrated lattice with longitudinal $t_x$,
transverse $t_y$, and diagonal $t_d$ hopping parameters.}
\label{lattice}
\end{figure}

We consider the Hubbard model with the local interaction $U$ and
 the following non-interacting single-particle energies,

\begin{eqnarray}
\nonumber  \epsilon(k_x,k_y)=-2t_xcosk_x-2t_ycosk_y-\\
2t_dcos(k_x+k_y)-2t_dcos(k_x-k_y),
\end{eqnarray}

\noindent the hopping parameters $t_x$, $t_y$, and $t_d$, respectively 
in the longitudinal, transverse, and diagonal directions, are illustrated in
Fig.\ref{lattice}. The presence of $t_d$ ensures that the non-interacting 
Fermi surface is not perfectly nested. $t_y$ and $t_d$ must be 
$(t_y,t_d) \ll t_x$ for the two-step DMRG method to be accurate. In this study, 
we set $t_x=1$ and $t_y=t_d=0.05t_x$. The choice of this model thus precludes 
the theoretical possibility of the nesting induced exponentially small gap.
The band-width is $W=4.4 t_x$, we set $u=U/W$. 

\section{Two-step density-matrix renormalization group}
\label{tsdmrg}

The two-step DMRG is a generalization of the conventional DMRG method 
\cite{white} to quasi-1D Hamiltonians. The DMRG is a RG procedure in which 
the reduced density-matrix is used to retain the most important states of the 
system. The DMRG itself is a crucial improvement over the block RG method 
\cite{slac} which extended the Wilson RG method \cite{wilson} used in the 
solution of the Kondo impurity problem to lattice models. The
block method has a major handicap, by dividing the lattice into
independent blocks, it neglects at its initial step the inter-block
interaction. But if the inter-block interaction is of the same order
as the intra-block interaction, this introduces an error from which it
is difficult to recover even by keeping a large number of states. In
the DMRG the lattice is built by initially coupling the block to the
rest of the lattice. Let us consider a system (S) coupled to an 
environment (E), let $N_s$ and $N_e$ be respectively the number of 
states respectively of the system and for the environment. Let $\Phi$ be 
for instance the ground-state wave function of the super-system including 
the system and the environment, 

\begin{eqnarray}
\Phi(S,E)=\sum_{i_s=1,N_s;i_e=1,N_e}\alpha_{i_s,i_e}\psi_{i_s}\chi_{i_e},
\end{eqnarray}

\noindent where the $\psi_{i_s}$'s represent the system's basis states and
the $\chi_{i_e}$'s the environment basis states; $N_s$ and $N_e$ are
respectively the total number of states of the system and of the 
environment.  The essence of the RG procedure is the truncation of the
Hilbert's space, starting with a small system for which the total number
of states can be kept, at some step when the lattice gets large, only 
a smaller number $m_s < N_s$ of the system's states can be kept. The
error in this truncation is given by the eigenvalues $\lambda_{i_s}$ of the 
reduced density-matrix of the system,

\begin{eqnarray}
D_S=\sum_{i_e=1,N_e}\Phi(S,E)\Phi^*(S,E).
\end{eqnarray}

\noindent From the relation,

\begin{eqnarray}
\sum_{i_s=1,N_s} \lambda_{i_s}=1,
\end{eqnarray}

\noindent the error made by representing the system by $m_s$ states instead of $N_s$
is given by,

\begin{eqnarray}
\rho=1-\sum_{i_s=1,m_s} \lambda_{i_s}.
\end{eqnarray}

For a large number of 1D models, $\rho$ is very small if $m_s$ is only a
few hundreds. Application of the DMRG method to Heisenberg chains
with $S=1/2$ or $S=1$ \cite{white}, $m_s \alt 100$, the ground-state energy, 
correlation functions and lowest excitation gap were obtained with an 
astonishing accuracy. 

It was hoped that, given the level of accuracy of the DMRG 
for 1D models, the method would also perform reasonably well for 2D models. 
 However, for a 2D lattice, the value of $m_s$ necessary
to retain good accuracy appears to increase exponentially with the system
size. This is related to the entropy area law which predicts an exponential 
increase of $m_s \propto 2^{L^{D-1}}$ in 2D. The entropy area law implies that the
direct application of the 2D DMRG would only be limited to relatively
narrow systems, it however leaves a window of success for quasi-1D 
systems as we will explain below. The study of quasi-1D models would
yield valuable information about the corresponding isotropic models.
 Most importantly the two-step approach had a direct relevance 
to the physical properties of quasi-1D materials for which 
$t_y \ll t_x$ such as the organic and inorganic quasi-1D conductors.

Let us consider for instance the Hubbard chain with a charge gap $\Delta$. 
If the transverse coupling $t_y$ is infinitely small with respect to
$\Delta$, so that the system remains in the same phase as the decoupled
chains. It is obvious that the decoupled chain limit is a good starting point
to describe the weakly-coupled chain system. As $t_y$ increases, the quality
of decoupled chain as a starting point will decrease, if the same number
of states is kept, until $t_y$ reaches a 
quantum critical point $t_y^c$ at which the systems enters in the 2D regime. 
In principle, when $t_y$ is in the 2D phase, it would be wrong to start from 
the decoupled chain limit. This is because there are a huge number of low-lying
states with nearly equal weight in the reduced density-matrix. 

The important point which nevertheless makes calculations possible is that 
actual calculations are done on finite systems which
have a discrete spectrum. Thus even if $t_y$ has a value corresponding to 
the 2D phase for a system size $L$, given the discreteness of the energy 
spectrum for a finite system, if the energy width of the states kept is such 
that $\Delta E \gg t_y$, starting from decoupled chain might still lead to 
accurate results. For such a system, the DMRG can be used to study the 
ground-state phase transition since it will display a different scaling 
behavior above and below $t_y^c$. The same type of analysis may be used for 
gap-less chains as well, $\Delta(L)$ will yield the relevant energy scale above 
and below the transition.

The separation of the energy scales is basic idea of the two-step DMRG 
\cite{moukouri}. The two-step DMRG uses the 
extraordinary accuracy that the DMRG can achieve in 1D in two steps. In the 
first step, the low-energy Hamiltonian is obtained accurately using the DMRG. 
Then, in the next step  small transverse perturbations are inserted.  
The 2D effective Hamiltonian is 1D, the DMRG is again applied to solve the 
problem in the transverse direction. Indeed, this procedure is valid only if 
the transverse couplings are very small with respect to the longitudinal 
couplings.  The success of the two-step DMRG in yielding reliable results on 
the eventual new physics induced by the perturbation will depend on the value 
of the critical transverse coupling necessary to drive the systems in a new 
phase. If the magnitude of the perturbation $t_y$ necessary to drive the 
system away from the 1D physics is small in comparison with the width of the 
states kept, the two-step DMRG is expected to  be successful. This is for 
instance the case of coupled Haldane chains studied in section \ref{fsc}. 
However, if the magnitude of the perturbation is too  large, the two-step 
DMRG would not be able to describe the 2D physics accurately.

The real challenge in the two-step starts after finishing making the program 
code work. The essential part of the subsequent activity is finding a region 
in the parameter space of a given model where interesting 
physical results can be extracted.  For more details
about the two-step DMRG, we refer the reader to Ref.\cite{moukouri}.

\begin{table*}
\begin{ruledtabular}
\begin{tabular}{|c|c|c|c|c|c|c|}
 &$12 \times 13$&$16 \times 17$&$20 \times 21$&$24 \times 25$&$28 \times 29$& $32 \times 33$\\\hline
$\Delta E (u=0)$&$1.6220$&$1.2683$&$1.0410$&$0.8819$&$0.7685$&$0.6772$ \\\hline
$\rho_1 (u=0)$&$8\times10^{-9}$&$3\times10^{-7}$&$7\times10^{-7}$&$1\times10^{-6}$&$3\times10^{-6}$&$4\times10^{-6}$ \\\hline
$\rho_2 (u=0)$&$0$&$0$&$0$&$0$&$0$&$4\times10^{-4}$ \\\hline
$\Delta E (u=0.4261)$&$1.5630$&$1.2333$&$1.0204$&$0.8733$&$0.7825$&$---$ \\\hline
$\rho_1 (u=0.4261)$&$1\times10^{-7}$&$3\times10^{-7}$&$7\times10^{-7}$&$1\times10^{-6}$&$2\times10^{-6}$&$---$ \\\hline
$\rho_2 (u=0.4261)$&$2\times10^{-8}$&$1\times10^{-7}$&$3\times10^{-7}$&$5\times10^{-7}$&$2\times10^{-6}$&$---$ \\\hline
$\Delta E (u=0.6818)$&$1.6249$&$1.3121$&$1.1128$&$0.9907$&$0.9134$&$0.8463$\\\hline
$\rho_1 (u=0.6818)$&$9\times10^{-8}$&$3\times10^{-7}$&$5\times10^{-7}$&$1\times10^{-6}$&$2\times10^{-6}$&$3\times10^{-6}$ \\\hline
$\rho_2 (u=0.6818)$&$2\times10^{-8}$&$8\times10^{-8}$&$1\times10^{-7}$&$2\times10^{-7}$&$2\times10^{-7}$&$7\times10^{-7}$  
\end{tabular}
\end{ruledtabular}
\caption{Energy width $\Delta E$, truncation errors $\rho_1$ (first DMRG step),  $\rho_2$ (second DMRG step) for $u=0$, $u=0.4261$ (near the quantum critical
point), and for $u=0.6818$ in the Hubbard lattice when $m_1=512$ and  $m_2=96$ 
states are retained.} 
\label{energy-width}
\end{table*}

In the first step of the DMRG, we targeted charge sectors with 
$N_e, N_e\pm 1, N_e \pm 2$, where $N_e$ corresponds to the number of
electrons at half-filling; for each charge sector, we targeted
the spin sectors with the lowest $S_z, S_z \pm 1$; hence we targeted
a total of $n_{targ}=17$ charge-spin sectors during each DMRG iteration.
The reduced density-matrix was given by,

\begin{eqnarray}
D_S=\sum_{k=1,n_{targ}} \omega_k\sum_{i_e=1,N_e}\Phi_k(S,E)\Phi_k^*(S,E).
\end{eqnarray}

\noindent where we assigned an equal weight $\omega_k=1/17$ to each
state $\Phi_k$. In all the simulations we kept $ms_1=512$ states such that 
the largest truncation error was $\rho_1 \approx 10^{-6}$ for systems of up to 
$L_x=32$ as can be seen in Table \ref{energy-width}. 

In the second step, we targeted $n_{targ}=3$ charge sectors $N_e, N_e \pm 1$ 
with the lowest $S_z$. The reduced density-matrix was formed by attributing an 
equal weight $\omega_k=1/3$ for each of $k=1,n_{targ}$ states. We kept $ms_2 =96$ 
states such that the width of the retained states,
$\Delta E \gg t_y,t_d$ for $t_d=t_y=0.05t_x$. $\Delta E$ is displayed in
Table \ref{energy-width}. For these parameters, the truncation error
during the second step was such that $\rho_2 \alt \rho_1$ for systems of
up to $L_x \times L_y=32 \times 33$ when three superblock states were
targeted. We empirically chose $ms_2$ such that $\Delta E/t_y=10$. 
For this ratio, we can accurately reproduce the exact result at $u=0$.

\section{Finite-size scaling}
\label{fsc}

\subsection{General concepts}

\begin{figure}
\begin{center}
\includegraphics[width=6.cm, height=4cm]{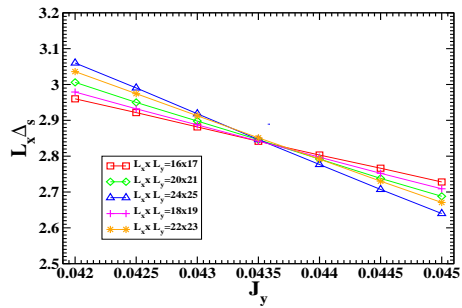}
\end{center}
\caption{ Scaled spin gap in the quasi-1D Heisenberg model as function
of $J_y$.}
\label{sgap1}
\end{figure}

\begin{figure}
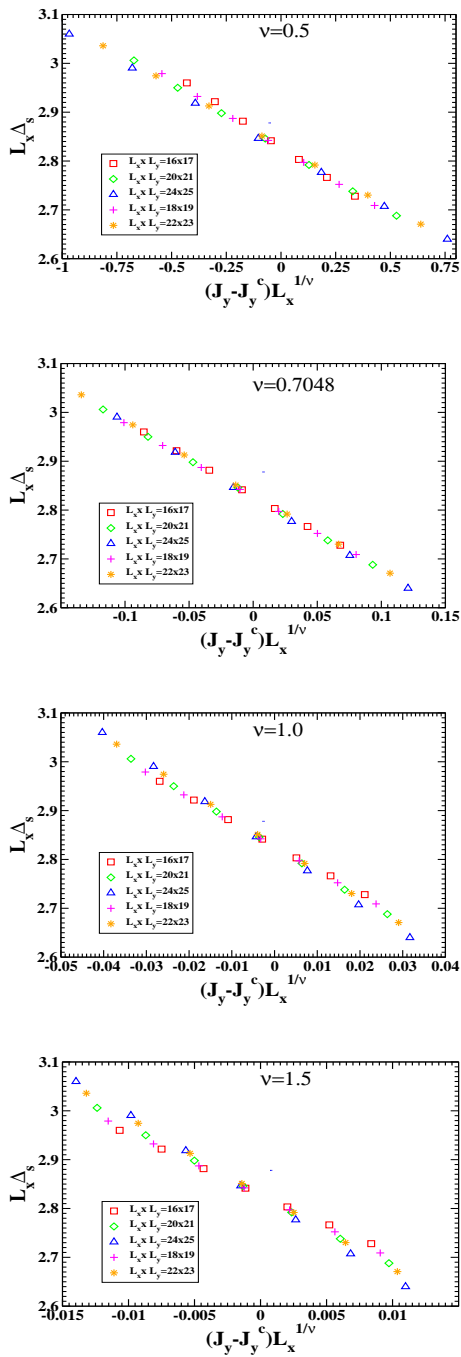

\begin{center}
$\begin{array}{c@{\hspace{0.25in}}c}
\vspace{0.5cm}
\includegraphics[width=6.cm, height=4cm]{gapgp0.042-0.045_nu0.5.eps}
\end{array}$
$\begin{array}{c@{\hspace{0.25in}}c}
\vspace{0.5cm}
\includegraphics[width=6.cm, height=4cm]{gapgp0.042-0.045_nu0.7.eps}
\end{array}$
$\begin{array}{c@{\hspace{0.25in}}c}
\vspace{0.5cm}
\includegraphics[width=6.cm, height=4cm]{gapgp0.042-0.045_nu1.0.eps}
\end{array}$
$\begin{array}{c@{\hspace{0.25in}}c}
\includegraphics[width=6.cm, height=4cm]{gapgp0.042-0.045_nu1.5.eps}
\end{array}$
\end{center}
\caption{ $\Delta \times L_x$ as function of $u$ ((a) and (b), as
function of $(J_y-J_y^c)L_x^{1/\nu}$ for different $L_x \times L_y$ and
for different universality classes: mean-field ($\nu=0.5$), classical
3D Heisenberg ($\nu=0.7048$), 2D Ising ($\nu=1.0$), fictitious class
($\nu=1.5$).}
\label{gapscoll}
\end{figure}

We apply finite-size scaling \cite{barber} to analyze the results on the 
 charge gap $\Delta$. The procedure is simple. We accurately compute 
$\Delta$ in order to locate the quantum critical point. We then collapse
the data using the exponents $\nu$ of known universality classes in order 
to find the class corresponding to the Mott transition. We emphasize that
in this procedure there is no extrapolation or external parameter besides
the data and the exponent of the chosen universality class.

The accurate location of the critical point is 
done by plotting the product $L_x^{-1} \xi$ as function of the interaction
driving the transition. $\xi$ is the correlation length. This is because at 
the transition, $L_x^{-1} \xi$ is independent of $L_x$. For the gap the 
function $L_x^{-1} \xi$ translates to $L_x^{-z} \Delta^{-1}$, where $z$ is 
the dynamical exponent. Near the the quantum critical point,  the product 
$L_x^z \Delta$ is given by a universal function,

\begin{eqnarray}
L_x^z\Delta=f((g-g_c)L_x^{1/\nu}),
\end{eqnarray}

\noindent where $g$ is a generic coupling driving the transition, $g_c$ is 
its magnitude at the quantum critical point, and $\nu$ is the correlation 
length critical exponent.

\subsection{Application to coupled Heisenberg chains with $S=1$}

In Fig.\ref{sgap1}, we illustrate the finite-size analysis that we apply
below to weakly coupled Heisenberg chains with $S=1$. The model which was 
studied in Ref.\cite{moukouri-eidelstein} is given by the Hamiltonian,

\begin{eqnarray}
H_s=J_x\sum_{i_x,i_y} {\bf S}_{i_x,i_y}{\bf S}_{i_x+1,i_y}+
J_y\sum_{i_x,i_y} {\bf S}_{i_x,i_y}{\bf S}_{i_x,i_y+1}.
\label{heisenberg}
\end{eqnarray}

In the model (\ref{heisenberg}), there is transition from a magnetically 
disordered ground state, the Haldane gap phase, to a magnetically ordered 
ground state which is induced by the transverse coupling $J_y$. This 
transition has been studied by the quantum Monte Carlo method 
\cite{matsumoto}. In this transition $z=1$, and it belongs to the 
universality class of the 3D classical Heisenberg model, for which 
$\nu=0.7048$ \cite{landaudp}. In Fig.\ref{sgap1} we plot $L_x\Delta_s$ 
as function of $J_y$, where $\Delta_s$ is the spin gap. We studied
systems ranging from $L_x \times L_y=12 \times 13$ to $24 \times 25$.
We applied periodic boundary conditions along the $x$-direction and
open boundary conditions along the y-direction. At the
quantum critical point $J_y=J_y^c$, $L_x\Delta_s$ is independent of $L_x$.
There are small size effects for smaller systems. We thus included
only systems larger than $16 \times 17$. All the curves $L_x \Delta_s$ cross 
at $J_y^c$. The critical point $J_y^c=0.04368$ was located graphically.
It is in perfect agreement with the quantum Monte Carlo value 
$J_y^c=0.043648(8)$. 

The determination of the universality class is done by plotting $L_x \Delta_s$
as function of $(J_y-J_y^c)L_x^{1/\nu}$. In Fig.\ref{gapscoll}, 
$L_x \Delta_s$ is displayed for different values of $\nu$ corresponding
to mean-field, classical 3D Heisenberg, 2D Ising, and a fictitious 
universality class with $\nu=1.5$. As expected from Monte Carlo simulations, 
the best data collapse was obtained
for $\nu \approx 0.7048$ which is predicted Monte Carlo value \cite{landaudp}
for the classical 3D Heisenberg universality class. 

\section{Results and discussion}
\label{results}

\subsection{Correlation length exponent at the Mott transition}

\begin{figure}
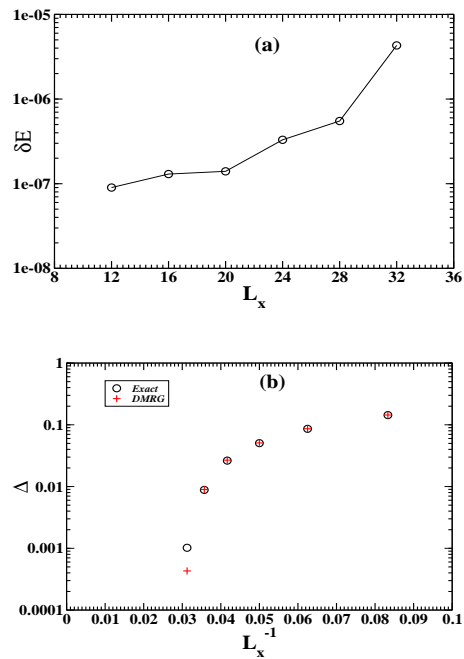

\begin{center}
$\begin{array}{c@{\hspace{0.25in}}c}
\vspace{0.5cm}
\includegraphics[width=6.cm, height=4cm]{gserror.eps}
\end{array}$
$\begin{array}{c@{\hspace{0.25in}}c}
\includegraphics[width=6.cm, height=4cm]{gapu0t0.05.eps}
\end{array}$
\end{center}
\caption{ Error in the ground-state energy for quasi-one-dimensional
systems as function of the linear dimension $L_x$ of the lattice.
Single-particle two-step DMRG gaps versus exact gaps
as function of $L_x$}
\label{gserror}
\end{figure}

 We can now confidently apply the same method to the Hubbard model. It 
has roughly the same level of difficulty as the coupled Heisenberg chain
problem. First, we compared the
two-step DMRG results with the exact energies at $u=0$. We emphasize that
this test is non-trivial for a real-space technique such as the DMRG 
because in real space, the hopping term is non-diagonal. 
In Fig.\ref{gserror}(a),
we show the error $\delta E$ in the ground-state energies per site for systems
ranging from $L_x \times L_y= 12 \times 13$ to $32 \times 33$. The two-step
DMRG is in very good agreement with the exact result; $\delta E < 10^{-6}$ 
and increases relatively slowly with $L_x$ for systems 
$L_x \times L_y < 28 \times 29$ and starts to grow sharply beyond this size.
In Fig.\ref{gserror}(b), we compare the single-particle gap,
$\Delta=\frac{1}{2}[E_0(N+1)+E_0(N-1)-2E_0(N)]$, obtained with the
two-step DMRG to the exact gap. The largest error for the gap was
about $5 \times 10^{-4}$ in the $32 \times 33$ systems. Since for this
size the exact gap is only $\Delta =0.00103$, we excluded the $32 \times 33$
systems from the data used to extract the critical exponent. For the largest 
systems kept for the analysis $28 \times 29$, the two-step DMRG gap is 
$\Delta=0.00895$ which is to be compared to the exact gap $\Delta=0.00883$. 
The relatively large loss of accuracy in the gap for $32 \times 33$ systems 
follows from the sharp increase in $\delta E$.

\begin{figure}
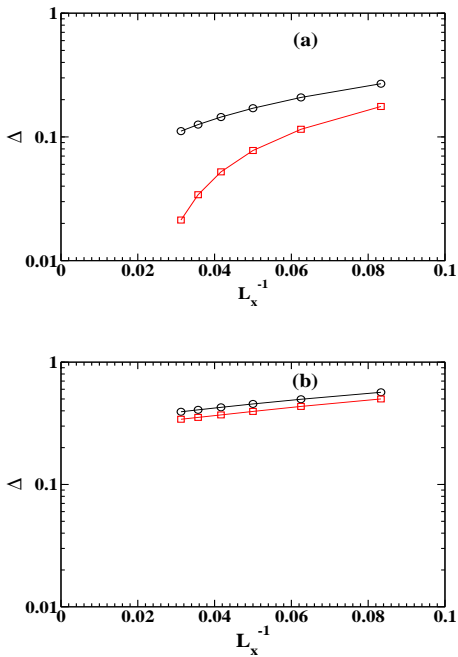

\begin{center}
$\begin{array}{c@{\hspace{0.25in}}c}
\vspace{0.5cm}
\includegraphics[width=6.cm, height=4cm]{gapu1.eps}
\end{array}$
$\begin{array}{c@{\hspace{0.25in}}c}
\includegraphics[width=6.cm, height=4cm]{gapu3.eps}
\end{array}$
\end{center}
\caption{Quasi-particle gaps as function of $L_x$ for two characteristic
values of the interaction: (a) $u=0.2273$, (b) $u=0.6818$ for 1D (circles) 
and quasi-1D (squares) systems.}
\label{gapu1-3}
\end{figure}

When $u \neq 0$, the two-step DMRG retains the same level of accuracy 
as at $u=0$. This is because, when the same number of states
$m_2$ is kept, the truncation error $\rho$ remains close to that of
$u=0$ as seen in Table \ref{energy-width}. $\Delta E$ slightly increases
with $u$, hence, the condition $\Delta E \gg t_y,t_d$ is also fulfilled.
Unlike the pure 1D model, the metallic phase is expected to have a finite
width in the quasi-1D model. In Fig. \ref{gapu1-3} we show the gap as function
of $L_x$ for two characteristic values of the interaction at $u=0.2273$ and
$u=0.6818$ for the 1D and quasi-1D systems. There appears to be two regimes.
In Fig.\ref{gapu1-3}(a), for $u=0.2273$ the quasi-1D gap shows a sharp decay in 
contrast to the 1D gap which decays more slowly. This is consistent with the 
finite value of the 1D gap and the presumably zero value of the quasi-1D gap
in the thermodynamic limit. In Fig.\ref{gapu1-3}(b), for $u=0.6818$ both gaps 
remain very close and have a finite value in the thermodynamic limit. This 
behavior suggests that there would be a quantum critical point at 
$0.2273 \alt u_c \alt 0.6818$. We would like to emphasize that in 
Ref.\cite{moukouri-eidelstein}, in 1D in agreement with the exact 
result\cite{lieb-wu} the DMRG yielded $u_c=0$. 

\begin{figure}
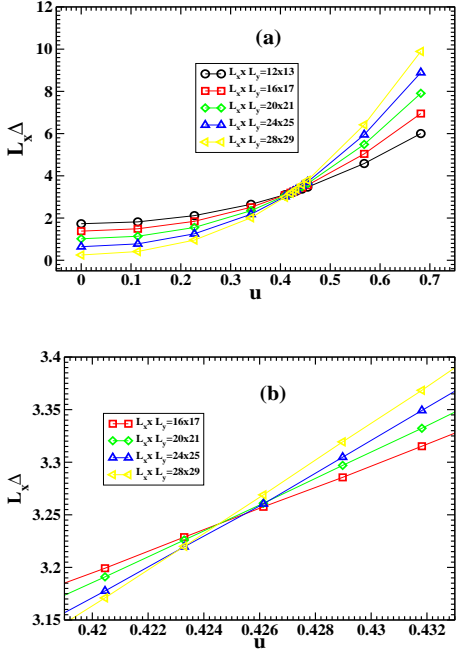

\begin{center}
$\begin{array}{c@{\hspace{0.25in}}c}
\vspace{0.5cm}
\includegraphics[width=6.cm, height=4cm]{gapt0.05_2d_2.eps}
\end{array}$
$\begin{array}{c@{\hspace{0.25in}}c}
\vspace{0.5cm}
\includegraphics[width=6.cm, height=4cm]{gapt0.05_2d_2b.eps}
\end{array}$
\end{center}
\caption{ $\Delta \times L_x$ as function of $u$ for the Hubbard
model: (a) extendend range of $u$, (b) for $u$ in the vicinity
of the quantum critical point.}
\label{gap3}
\end{figure}

We analyze our results using the language of second order transitions.
This is justified because we did not see any sharp change in our data
for the ground-state energy or the gap. Generally, in a first order transition
it would usually be expected that the ground-state energy would be 
non-differentiable and the gap would show a discontinuity at the transition 
point. These were not seen in our data.  The absence of a discontinuity is 
seen for instance in the behavior of $L_x \Delta$ in Fig.\ref{gap3}. This 
justifies the assumption that the transition is of second order.

As for the Heisenberg model above,  In Ref.\cite{capogrosso, hohenadler}, 
the value $z=1$ was predicted for the interaction induced Mott transition. 
But in the density induced transition the dynamical exponent is $z=2$. 
In order to find the value of $z$, we plotted both $L_x \Delta$ and
$L_x^2 \Delta$. However, the rough estimate of the critical value found for
$L_x^2 \Delta$, $u_c \approx 0.1705$ was very inconsistent with the direct
extrapolation of the data. For instance, at $u=0.2273$, $\Delta$ extrapolates
to $0$. This allows us to rule out $z=2$ as well as higher values of $z$
since they yield even smaller $u_c$.

We show for $z=1$, $L_x \Delta$ as function of $u$ in Fig.\ref{gap3}. A
first sweep of the interaction range $0 \le u \le 0.6818$ in Fig.\ref{gap3}(a)
indicates that $0.4 \le u_c \le 0.5$. In Fig.\ref{gap3}(b), to precisely
locate $u_c$, we concentrate in the interaction range $0.420 \le u \le 0.432$,
a graphical estimate yields $u_c=0.4255$. The range of values of $u$ for
the critical analysis $\delta u=0.02656 u_c$ is comparable to that used
in Ref.\cite{capogrosso} $|\delta (J/U)|=0.01526 (J/U)_c$ for the Bose Hubbard
model, and in Ref.\cite{hohenadler} $|\delta (t/g)|=0.01339 (t/g)_c$ for the 
Jaynes-Cummings-Hubbard model. $(J/U)$ and $t/g$ are the ratio of the hopping
parameter over the interaction. 

\begin{figure}
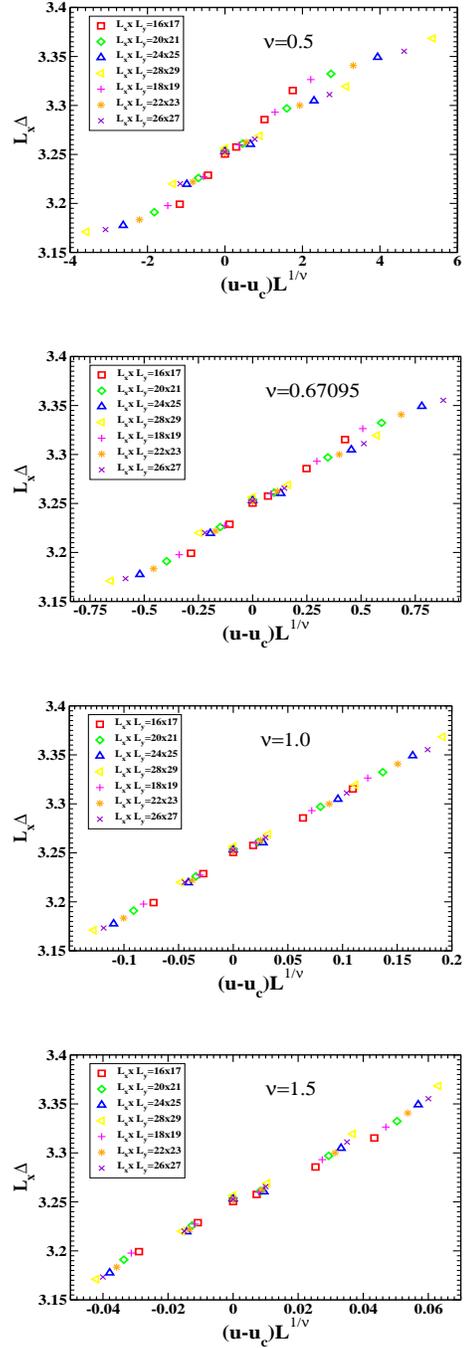

\begin{center}
$\begin{array}{c@{\hspace{0.25in}}c}
\vspace{0.5cm}
\includegraphics[width=6.cm, height=4cm]{gapcollapse_nu0.5.eps}
\end{array}$
$\begin{array}{c@{\hspace{0.25in}}c}
\vspace{0.5cm}
\includegraphics[width=6.cm, height=4cm]{gapcollapse_nu0.6709.eps}
\end{array}$
$\begin{array}{c@{\hspace{0.25in}}c}
\vspace{0.5cm}
\includegraphics[width=6.cm, height=4cm]{gapcollapse_nu1.0.eps}
\end{array}$
$\begin{array}{c@{\hspace{0.25in}}c}
\includegraphics[width=6.cm, height=4cm]{gapcollapse_nu1.5.eps}
\end{array}$
\end{center}
\caption{ $\Delta \times L_x$ as function of $u$ ((a) and (b), as
function of $(u-u_c)L_x^{1/\nu}$ for different $L_x \times L_y$ 
for $\nu$ corresponding to different universality classes:
mean-field ($\nu=0.5$), 3D classical XY ($\nu=0.67095$),
2D Ising ($\nu=1.0$), fictitious ($\nu=1.5$).}
\label{gapcoll}
\end{figure}

As for the Heisenberg model above, we determine the universality
class of the Hubbard model by plotting $L_x \Delta$ as function of 
$(u-u_c)L^{1/\nu}$. In  Fig.\ref{gapcoll}. We tried different values of 
$\nu$ corresponding to the mean-field $\nu=0.5$, 3D XY, 2D Ising $\nu=1.0$, 
and a fictitious $\nu=1.5$ cases. For the 3D
XY model, Monte Carlo values $\nu$ are found between
 $\nu=0.662(7)$ and $\nu=0.6723$ \cite{hasenbusch}, and with the bosonic 
Hubbard model \cite{capogrosso} and the Jaynes-Cummings-Hubbard model
 \cite{hohenadler} for which $\nu=0.6715$. The experiments on $H_e^4$ films
are believed to yield the best estimate of $\nu$ for the 3D XY models. 
Experiments have smaller errors than Monte Carlo simulations.
For instance $\nu$ was found to be $\nu=0.6708(4)$ in Ref.\cite{swanson}, 
$\nu=0.6705(6)$ in Ref.\cite{goldner}, and $\nu=0.67095(13)$ in
Ref.\cite{lipa}. We used this last value to collapse the data for the
test of the 3D XY universality class. 

Fig.\ref{gapcoll},  clearly shows that the best fit to the data is obtained
for $\nu=1.0$. This implies that the Mott transition in the Hubbard model
belongs to the universality class of the 2D Ising model as predicted
by the approximate mapping of Ref.\cite{castellani}.

The 3D XY universality class for the Mott transition in 2D was  
conjectured in approximate slave-rotor analyses of the 
fermionic Hubbard model in Ref.\cite{florens-georges,senthil}. 
This work shows that the neglect of the gauge field 
during the factorization of the fermionic operators into a
spinless boson and a charge-less spin is not justified. It should
be noted that the 3D Ising and 3D Heisenberg universality class
for which $\nu$ is close to that of the 3D XY class, 
respectively $\nu=0.6298(5)$ \cite{hasenbusch2}, $\nu=0.7048$ \cite{landaudp}
 were also ruled out.  

\subsection{Order parameter for the Mott transition}

The identification of the universality class of the Mott transition
suggests the following analogy with the Ising transition. The weak
$u$ limit should correspond to the high temperature phase in the
Ising model. At $u=0$, the four possible local states, $|0 \rangle$,
$|\uparrow \rangle$, $|\downarrow \rangle$, and $|\uparrow \downarrow \rangle$
are equally probable respectively with

\begin{equation} 
n_0= {\tilde n}_{\uparrow}={\tilde n}_{\downarrow}= n_d=\frac{1}{4},
\end{equation}

\noindent in ${\tilde n}_{\uparrow}$ and ${\tilde n}_{\downarrow}$ only purely
singly occupied sites are counted,

\begin{eqnarray}
{\tilde n}_{\uparrow}=n_{\uparrow}-n_d,\\
{\tilde n}_{\downarrow}=n_{\downarrow}-n_d.
\end{eqnarray}

In the opposite limit $u=\infty$ which corresponds to the low temperature
phase, holes and doubly occupied sites are not allowed, 

\begin{eqnarray}
n_0=n_d=0,\\ 
{\tilde n}_{\uparrow}={\tilde n}_{\downarrow}=\frac{1}{2},
\end{eqnarray}
 
\noindent the local possible states have shrinked from 4 to 2 due to the 
 $Z_2$ Ising  symmetry breaking. This is in
contrast to the slave-rotor analyses  where the Fermi liquid is regarded
as the ordered phase.
 The isomorphism $SU(2)/Z_2 \equiv SO(3)$ implies
that in principle after the Mott transition,  the effective
spin Hamiltonian, obtained by projecting out the empty and doubly
occupied states,  should retain the full spin rotational symmetry. The
eventual spin long-range order will depend on the couplings present in the 
effective Hamiltonian.

The natural order parameter $M$ for the Mott transition should thus be given 
by the average number of singly occupied sites minus the number of doubly 
occupied and empty sites,

\begin{equation}
M=\langle {\tilde n}_{\uparrow}+{\tilde n}_{\downarrow}-n_d-n_0 \rangle.
\end{equation}
 
\noindent Thus,

\begin{equation}
M=\langle n-4n_d \rangle.
\end{equation}

\begin{figure}[b]
\begin{center}
\vspace{0.5cm}
\includegraphics[width=6.cm, height=4cm]{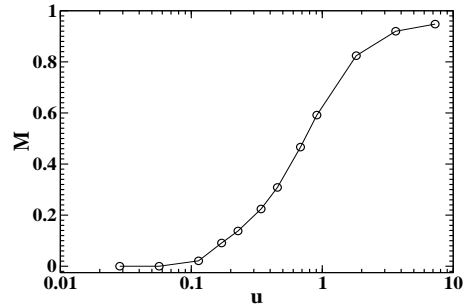}
\end{center}
\caption{Order parameter $M$ of the Mott transition as function of $u$ for
a $24 \times 25$ system.}
\label{orderpar}
\end{figure}

For $u=0$, $M=0$ and for $u=\infty$, $M=1$. It should be expected
that for $u \leq u_c$, $M=0$. But this is not true for finite systems.
Because of the finite size gap, finite systems are always 'ordered', thus
$M$ will always have a finite value for a finite system even when
$u \leq u_c$. Since $n_d$ is a local quantity, it changes very slowly
with system sizes. This means that very large systems are necessary to 
extrapolate accurately to its thermodynamic value. In the two-step
DMRG approach , it is  more judicious to calculate the correlator,

\begin{equation}
M=\frac{1}{L_x}\sqrt{\langle\sum_i M_0M_i \rangle},
\end{equation}

\noindent for the middle chain. $M$ is shown in Fig.\ref{orderpar} for a 
$24 \times 25$ system. The curve of $M$ has the usual form of an order 
parameter curve. However in the vicinity of the quantum critical point, 
because of the use of open boundary conditions, the data are 
strongly affected by the 2D remnant of Friedel
oscillations. Convergence is very slow even with this definition of $M$.
It can be seen that the value of $M$ is still appreciable at the quantum
critical point $u=0.4255$. Significantly more work will be necessary
in order to reliably extract the order parameter exponent $\beta$.

\section{Conclusion}

In this paper, we used the two-step DMRG to analyze the finite size behavior
of the quasi-particle gap in the ground-state Mott transition in the quasi-1D 
Hubbard model. We chose a non-bipartite lattice to avoid the issue related
to the possible nesting induced Slater transition. We studied systems
ranging from $12 \times 13$ to $32 \times 33$. We were able to find the
universality class of the Mott transition in an un-biased calculation. 

 In contrast to the pure 1D model, we find that the quasi-1D models displays 
a genuine Mott transition at a finite critical interaction. Moreover, the
quasi-1D solution does not have the pathologies of the infinite dimensional
solution. It could thus serve as a basis for more realistic studies of the
detailed and well controlled analysis of the Mott transition. The critical 
behavior of the quasi-1D model Hubbard model is found to belong to the 
universality class of the 2D Ising model. The fact that 
the transitions in the quasi-1D Heisenberg and Hubbard models belong to the 
universality classes of their isotropic counterparts shows 
that despite the restriction of the two-step DMRG method to highly 
anisotropic 2D models, it is nevertheless very useful for the understanding 
of the physics of isotropic 2D systems.

 We did not  discuss the spin degrees of freedom. They are expected to be 
gap-less in either side of the Mott transition. In the insulating phase, in 
the strong coupling limit $U \gg t_x,t_y,t_d$, the anisotropic frustrated 
Hubbard model is equivalent to the anisotropic $J_1-J_2$ model with 
$J_x=t_x^2/U$, $J_y=t_y^2/U$, and $J_d=t_d^2/U$, where $J_x$, $J_y$, and 
$J_d$ are respectively the exchange parameter in the longitudinal, transverse,
and diagonal directions. Our choice $t_y=t_d$ implies that the ground state
will be magnetically ordered with the momentum ${\bf q}=(\pi,0)$. For
intermediate $U$, in the Mott insulator phase,  double occupation
 is not negligible straighforward mapping to the Heisenberg model
is not valid. However, the charge gap opening implies spin-charge
separation. Thus even in this case, the effective low-energy Hamiltonian 
should be Heisenberg like, albeit with non-trivial exchange parameters. 
Magnetic long-range order should be expected. However, a gap-less 
spin-liquid ground state with a spinon Fermi surface as 
suggested in Ref.\cite{florens-georges,senthil} is also possible.    

\begin{acknowledgments}
This work was supported in part by a Shapira fellowship of the Israeli Ministry
of Immigrant Absorption (S.M.), and by the Israel Science Foundation through 
grant no. 1524/07.
\end{acknowledgments}


\begin{thebibliography}{99}
\bibitem{mott} N.F. Mott, Proc. Phys. Soc. (London) {\bf A62}, 416 (1949).
\bibitem{imada-RMP} M. Imada, A. Fujimori, Y. Tokura, Rev. Mod. Phys. 
                   {\bf 70}, 1039 (1998).
\bibitem{hubbard} J. Hubbard, Proc. Roy. Soc.(London) {\bf A277}, 237 (1964).
\bibitem{lieb-wu} E.H. Lieb and F.Y. Wu, Phys. Rev. Lett. {\bf 20}, 1445 
                  (1968).
\bibitem{metzner-vollhardt} W. Metzner and D. Vollhardt, Phys. Rev. Lett. 
                          {\bf 62}, 324 (1989).
\bibitem{georges-kotliar} A. Georges and G. Kotliar, Phys. Rev. {\bf B 45},
                          6479 (1992).
\bibitem{georges-RMP} A. Georges, G. Kotliar, W. Krauth, and M.J. Rozenberg,
                      Rev. Mod. Phys. {\bf 68}, 13 (1996).

\bibitem{moukouri-jarrell} S. Moukouri and M. Jarrell, Phys. Rev. Lett. 
                          {\bf 87}, 167010 (2001).
\bibitem{parcollet} O. Parcollet, G. Biroli, G. Kotliar, Phys. Rev. Lett.
                          {\bf 92}, 226402 (2004).
\bibitem{maier-RMP} T. Maier, M. Jarrell, T. Pruschke, M. H. Hettler,
                    Rev. Mod. Phys. {\bf 77}, 1027 (2005).
\bibitem{zhang-imada} Y.Z. Zhang and M. Imada, Phys. Rev. {\bf B 76},
                           045108 (2007).
\bibitem{ohashi} T. Ohashi, T. Momoi, H. Tsunetsugu, and N. Kawakami,
                Phys. Rev. Lett. {\bf 100}, 076402 (2008).
\bibitem{park-haule} H. Park, K. Haule, and G. Kotliar, Phys. Rev. Lett.,
                           {\bf 101}, 186403 (2008).
\bibitem{potthoff} N. Balzer, B. Kyung, D. S\'en\'echal, A.-M.S. Tremblay,
                   and M. Potthoff, Eur. Phys. Lett. {\bf 85},
                   17002 (2009).
\bibitem{fisher} M. P. A. Fisher, P. B. Weichman, G. Grinstein, and D. S.
                 Fisher, Phys. Rev. {\bf 40}, 546 (1989).
\bibitem{elstner} N. Elstner and H. Monien, Phys. Rev. {\bf 59},
                   12184 (1999).
\bibitem{capogrosso} B. Capogrosso-Sansone, S. Gunes Soyler, N. Prokof'ev,
                     and B. Svistunov, Phys. Rev. {\bf A 77}, 015602 (2008). 
\bibitem{hohenadler} M. Hohenadler, M. Aichhorn, S. Schmidt, and L. Pollet,
                    Phys. Rev. {\bf A 84}, 041608 (R) (2011).
\bibitem{hirsch} J.E. Hirsch, Phys. Rev. {\bf B 31}, 4403 (1984).
\bibitem{varney} C.N Varney, C.-R. Lee, Z.J. Bai, S. Chiesa, M. Jarrell, and
                 R.T. Scalettar, Phys. Rev. {\bf B 80}, 075116 (2009).
\bibitem{florens-georges} S. Florens and A. Georges, Phys. Rev. {\bf B 70},
                           035114 (2004).
\bibitem{senthil} T. Senthil, Phys. Rev. {\bf 78}, 045109 (2008).
\bibitem{castellani} C. Castellani, C. Di Castro, D. Feinberg, and
                 J. Ranninger, Phys. Rev. Lett. {\bf 43}, 1957 (1979).
\bibitem{blume} M. Blume, V.J. Emery, and R.B. Griffiths, Phys. Rev.
                {\bf A 4}, 1071 (1971).
\bibitem{moukouri-eidelstein} S. Moukouri and E. Eidelstein, Phys. Rev.
                             {\bf 84}, 193103 (2011).

\bibitem{moukouri} S. Moukouri, Phys. Rev. {\bf B 70}, 014403 (2004).
\bibitem{bickers} N.E. Bickers and S.R. White, Phys. Rev. {\bf B 43},
                  8044 (1991).
\bibitem{matsumoto} M. Matsumoto, C. Yasuda, S. Todo, and H. Takayama,
                           Phys. Rev. {\bf 65}, 014407 (2001).
\bibitem{white} S.R. White, Phys. Rev. Lett. {\bf 69}, 2863 (1992).
\bibitem{slac} S. Drell, M Weinstein, and S. Yankielowicz, Phys. Rev. 
              {\bf D 14}, 487 (1976).
\bibitem{wilson} K.G. Wilson Rev. Mod. Phys.{\bf 47}, 773 (1975).

\bibitem{barber} M.N. Barber in 'Phase Transitions and Critical Phenomena', 
                 edited by C. Domb and J. L. Lebowitz, Academic Press, London,
                 Vol. 8, p. 145 (1983).

\bibitem{landaudp} K. Chen, A.M. Ferrenberg, and D.P. Landau, Phys. Rev.
                   {\bf 48}, 3249 (1993).
\bibitem{hasenbusch} M. Hasenbusch and T. T\"or\"ok, J. Phys. {\bf A32},
                      6361 (1999).
\bibitem{swanson} D.R. Swanson, T.C.P. Chui, and J. A. Lipa, Phys. Rev.
                  {\bf 46}, 9043 (1992).
\bibitem{goldner} L.S. Goldner, N. Mulder, and G. Ahlers, J. Low Temp. Phys.
                  {\bf 93}, 131 (1993).
\bibitem{lipa} J.A. Lipa, D.R. Swanson, J. Nissen, T.C.P. Chui, and
              U.E. Israelson, Phys. Rev. Lett. {\bf 76}, 944 (1996).
\bibitem{hasenbusch2} M. Hasenbusch, K. Pinn, S. Vinti, Phys. Rev.
                     {\bf B 59}, 11471 (1999).
\end{thebibliography}
\end{document}